\documentstyle[11pt,newpasp,twoside]{article}
\def\edcomment#1{\iffalse\marginpar{\raggedright\sl#1\/}\else\relax\fi}
\reversemarginpar

\begin{document}
\title{Between Cool Stars and Hot Planets: Origins of Brown Dwarfs}
 \author{Ray Jayawardhana}
\affil{Department of Astronomy, University of Michigan,\\
830 Dennison Building, Ann Arbor, MI 48109, U.S.A.}

\begin{abstract}
Brown dwarfs, which straddle the mass range between stars and planets,  
appear to be common both in the solar neighborhood and in star-forming 
regions. Their ubiquity makes the question of their origin an important 
one both for our understanding of brown dwarfs themselves as well as for
theories on the formation of stars and planets. Studies of young sub-stellar
objects could provide valuable insight into their formation and early 
evolution. Here I report on the 
latest results from our observational programs at Keck, VLT and Magellan 
on the disk and accretion properties of young brown dwarfs. We find 
compelling evidence that they undergo a T Tauri phase analogous to that 
of their stellar counterparts. 

\end{abstract}

\section{Introduction}

The past several years have seen the identification of a large number of 
sub-stellar objects in the solar neighborhood and in star-forming regions.
Yet their origin remains a mystery. One possibility is that they form like 
stars do as a result of the turbulent fragmentation and collapse of molecular 
cloud cores (e.g., Padoan \& Nordlund 2003). Another scenario which has 
gained popularity in recent times is that brown dwarfs are stellar embryos 
ejected from multiple proto-stellar systems (Reipurth \& Clarke 2001; 
Bate, Bonnell \& Bromm 2002).

There are few observational constraints on the formation and early evolution
of sub-stellar objects. Since studies of {\it young} brown dwarfs could 
provide valuable clues to their origin(s), we have commenced a multi-faceted
program to investigate the physical properties of brown dwarfs in 
star-forming regions and compare them to the much better studied low-mass
pre-main sequence stars. We have employed many of the methods developed 
in the study of 
T Tauri stars to address the key question of {\it whether young sub-stellar 
objects undergo a T Tauri-like phase.} 

\section{Disk Excess}
Using the ESO Very Large Telescope, Keck I and the NASA Infrared Telescope 
Facility, we have carried out a systematic study of infrared $L^\prime$-band 
(3.8$\mu$m) disk excess in a large sample of {\it spectroscopically 
confirmed} objects near and below the sub-stellar boundary in several 
nearby star-forming regions (Jayawardhana et al. 2003).

\begin{figure}
\plotfiddle{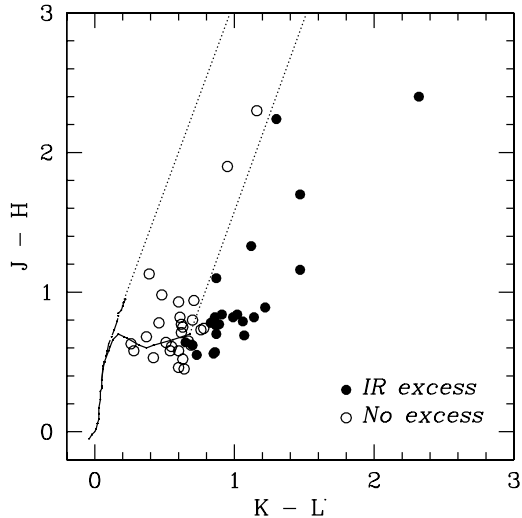}{2.0in}{0}{35}{35}{-100}{-100}
\vspace*{0.7in}
\caption{$J-H$/$K-L^\prime$ color-color diagram for our target sample. Also plotted are the empirical loci of colors for giants (solid) and for main-sequence dwarfs (dashed) from Bessell \& Brett (1988) and Leggett et al. (2002) and the reddening vectors (dotted). The filled circles are objects with  $E(K-L^\prime) >$ 0.2.}
\end{figure}

We find disk fractions of 40\%--60\% in IC 348, Chamaeleon I, Taurus and 
Upper Scorpius regions, using a conservative criterion of $K-L^\prime >$ 
0.2 for the presence of optically thick disks. ChaH$\alpha$ 2, which shows 
a large $K-L^\prime$ excess (0.97 mag) in our data is a probable close 
($\sim$0.2'') binary with roughly equal-mass companions (Neuh\"auser et al. 
2002). It is possible that a few of our targets harbor infrared companions 
that contribute to the measured excess, but this is unlikely in most cases. 
In IC 348, our disk fraction is comparable to that derived from H$\alpha$ 
accretion signatures in high-resolution optical spectra (Jayawardhana, 
Mohanty \& Basri 2003). However, in Taurus, Cha I and Upper Sco, which may 
be slightly older at $\sim$2-5 Myrs, we find $K-L^\prime$ excess in 
$\sim$50\% of the targets whereas accretion-like H$\alpha$ is seen in 
a smaller fraction of objects (Jayawardhana, Mohanty \& Basri 2002; 2003). 
This latter result suggests that dust 
disks may persist after accretion has ceased or been reduced to a trickle, 
as also suggested by Haisch, Lada \& Lada (2001). In the somewhat older 
($\sim$5 Myr) $\sigma$ Orionis cluster, only about a third of the targets 
show infrared excess. Neither of the two TW Hydrae brown dwarfs (age 
$\sim$ 10 Myr) in our sample shows excess. 

Our results, and those of Muench et al. (2001), Natta et al. (2002), and 
Liu, Najita \& Tokunaga (2003) show that a large fraction of very young brown 
dwarfs harbor near- and mid-infrared excesses consistent with dusty disks.
While the samples are still relatively small, the timescales for inner disk 
depletion do not appear to be vastly different between brown dwarfs and T 
Tauri stars. Far-infrared observations with the {\it Space InfraRed 
Telescope Facility} and/or the {\it Stratospheric Observatory For Infrared 
Astronomy} will be crucial for deriving the sizes of circum-sub-stellar 
disks and providing a more definitive test of whether brown dwarf disks
are truncated as predicted by the ejection scenario. 

\section{Accretion Signatures}
The shape and width of the H$\alpha$ emission profile is commonly used to 
discriminate between accretors and non-accretors among T Tauri stars (TTS). 
Stars exhibiting broad, asymmetric H$\alpha$ lines with equivalent width 
larger than 10 \AA~ are generally categorized as classical TTS (CTTS), 
although this threshold value varies with spectral type. We find that in 
very low mass (VLM) 
accretors, the H$\alpha$ profile may be somewhat narrower than that in 
higher mass stars. We propose that low accretion rates combined with 
small infall velocities at very low masses can conspire to produce this 
effect, and adopt $\sim$ 200 kms$^{-1}$ as a more appropriate, yet
conservative, threshold (see Jayawardhana, Mohanty \& Basri 2003 for
further discussion). 

\begin{figure}[hb]
\plotfiddle{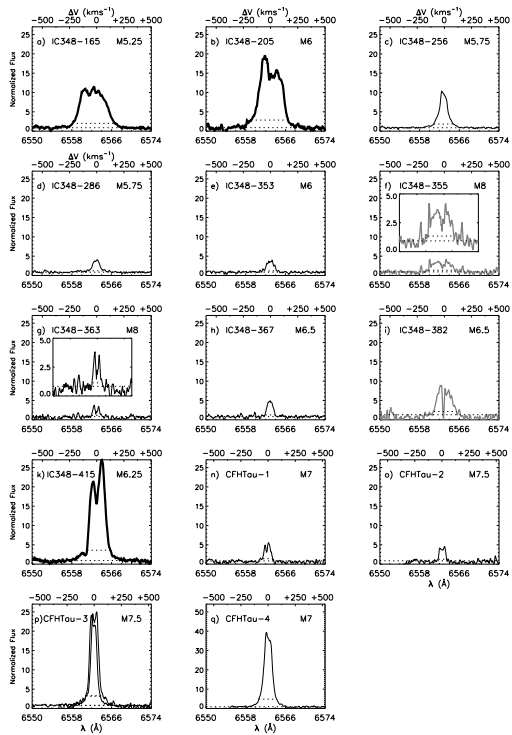}{2.4in}{0}{42}{42}{-100}{-90}
\vspace*{1.3in}
\caption{H$\alpha$ line profiles of IC 348 and Taurus targets. Thick black 
lines indicate accretors with broad H$\alpha$ as well as CaII and OI emission; 
grey indicates probable accretors, based on the H$\alpha$ profile-shape 
and 10\% full-width.} 
\end{figure}

We obtained high-resolution Keck and Magellan optical spectra of $\sim$45 
objects spanning the range of M5--M8 in several nearby star-forming regions. 
The vast majority of VLM objects in $\rho$ Ophiuchus  were inaccessible to 
our optical spectroscopy  because of significant extinction, presumably due 
to circumstellar as well as interstellar material. However, at least one of
our four $\rho$ Oph targets, GY 5, shows an accretion-like H$\alpha$ profile.
We also find evidence for accretion in 5/10 objects in IC 348 (Fig. 2), 
3/14 in Taurus, 1/12 in Cha I, and 1/11 Upper Scorpius (Jayawardhana, 
Mohanty \& Basri 2002, 2003). Perhaps somewhat suprisingly,
one of the three known brown dwarfs in the 10-Myr-old TW Hydrae association
also exhibits a broad, asymmetric H$\alpha$ line. If confirmed, this detection
suggests that accretion, albeit at very low rates, could last a fairly 
long time in some brown dwarfs (Mohanty, Jayawardhana \& Barrado y Navascu\'es
2003). 

\section{Evidence of an Outflow?}
We have recently obtained Magellan low-, medium-, and high-resolution 
optical spectra of a particularly intriguing young VLM object named 
LS-RCrA 1, identified by Fern\'andez and Comer\'on (2001). We confirm 
both pre-main sequence status and membership in the RCrA region for this 
object, through the detection of Li I, presence of narrow K I indicative 
of low gravity, and measurement of radial velocity.  The H$\alpha$ emission 
profile is very broad, with a 10\% full width of 316 kms$^{-1}$ at 
high-resolution, implying the presence of ongoing accretion.  Our 
spectra also exhibit many forbidden emission lines indicative of mass 
outflow (Fig. 3), in agreement with the Fern\'andez and Comer\'on results
(Barrado y Navascu\'es, Mohanty, \& Jayawardhana 2003). 

\begin{figure}[hb]
\plotfiddle{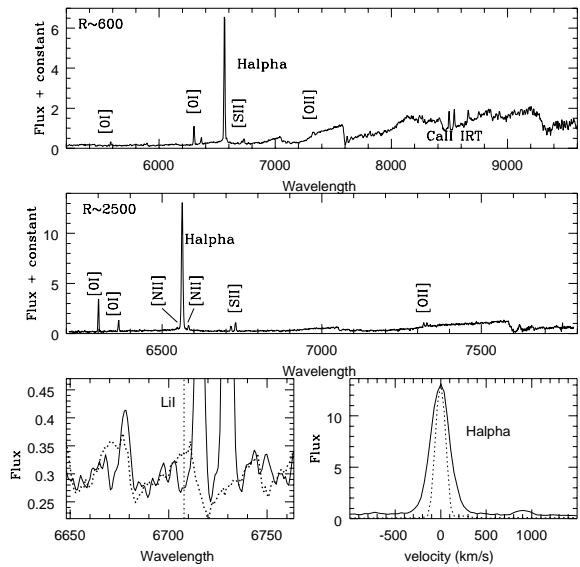}{2in}{0}{40}{40}{-120}{-120}
\vspace*{0.7in}
\caption{Spectral features of LS-RCrA 1. Upper panel.- Low resolution 
spectrum. Note the forbidden emission lines. Middle panel.- Medium 
resolution spectrum. Lower panel.- Zooms on Li I and H$\alpha$.}
\end{figure}

Our optical veiling measurements yield a mass accretion rate between 
10$^{-10}$ and 10$^{-9}$ solar masses per year. 
The presence of prominent outflow signatures at these low accretion 
rates is initially puzzling.  We consider, and discard as improbable, 
the possibility that these signatures arise in a line-of-sight Herbig-Haro 
knot unassociated with LS-RCrA~1 itself.  However, if LS-RCrA~1 possesses 
an edge-on disk, a natural outcome would be the enhancement of any outflow 
signatures relative to the photosphere; we favor this view. A low 
accretion/outflow rate, combined with an edge-on orientation, is further 
supported by the absence of high-velocity components and any significant 
asymmetries in the forbidden lines.  An edge-on geometry is also 
consistent with the lack of near-infrared excess in spite of ongoing 
accretion, 
and explains the relatively large H$\alpha$ 10\% width compared to other 
low-mass objects with similar accretion rates as well as the apparent
sub-luminosity of LS-RCrA~1 given its spectral type, distance and age. 
It would be extremely interesting to confirm the presence of an edge-on 
disk and a jet/outflow in this system with high-angular-resolution imaging. 

\section{Conclusion}
We have found compelling evidence, in the form of disk excesses and 
spectroscopic accretion signatures, that young brown dwarfs undergo
a T Tauri phase similar to that of solar-mass stars. In one case, there
is also a hint of possible mass outflow from a young sub-stellar object;
if confirmed, this would further strengthen the analogy with T Tauri
stars. 

\acknowledgments
I am most grateful to my collaborators in the study of young brown dwarfs: 
Subhanjoy Mohanty, Gibor Basri, Beate Stelzer, David Barrado y Navascu\'es, 
David Ardila, Karl Haisch, and Diane Paulson. This work was supported in 
part by the NSF grant AST-0205130.

\end{document}